\documentclass[aps,prd,twocolumn,showpacs,nofootinbib,floatfix]{revtex4-1}
\usepackage[english]{babel}
\usepackage{graphicx}
\usepackage{booktabs}
\usepackage{amssymb}
\usepackage{amsmath}
\usepackage{url}
\usepackage{cancel}
\usepackage{morefloats}
\usepackage{wrapfig}
\usepackage{array}
\usepackage[colorlinks,breaklinks]{hyperref}
\usepackage{preprintcover}
\usepackage{afterpage}
\usepackage[all]{nowidow}

\def\Pythia{{\sc Pythia}}
\def\BlackMax{{\sc BlackMax}}
\def\QBH{{\sc Qbh}}
\def\Geant{{\sc Geant4}}

\def\NLOJET{{\sc NLOJET++}}
\def\Herwig{{\sc HERWIG++}}
\def\BumpHunter{{\sc BumpHunter}}

\def\invpb{${\rm pb}^{-1}$}
\def\invfb{${\rm fb}^{-1}$}
\def\mjj{m_{jj}}

\def\qstar{{q^\ast}}
\def\Wstar{{W^\ast}}
\def\Wprime{W'}

\def\pT{$p_{\rm T}$}

\def\HighestDijetMass{4.5}
\def\integLumi{20.3~fb$^{-1}$}
\def\BHPvalue{0.075}
\def\BHZvalue{1.44}
\def\LowEdgeHighestBump{390}
\def\HighEdgeHighestBump{599}
\def\ChiSqPvalue{0.027}
\def\FitChiSquare{79} %
\def\BlindedFitChiSquare{37} %
\def\FitNDF{56}

\def\MjjResolutionLow{8\%}
\def\MjjResolutionHigh{4\%}
\def\MjjLimitqstar{4.06}
\def\MjjLimitExpectedqstar{3.98}
\def\MjjLimits8{2.70}
\def\MjjLimitExpecteds8{2.80}
\def\MjjLimitWstar{1.75}
\def\MjjLimitExpectedWstar{1.95}
\def\MjjLimitW1star{1.65}
\def\MjjLimitExpectedW1star{1.66}
\def\MjjLimitWprime{2.45}
\def\MjjLimitExpectedWprime{2.51}
\def\MjjLimitQBH{5.66}
\def\MjjLimitExpectedQBH{5.66}
\def\MjjLimitBlackMax{5.62}
\def\MjjLimitExpectedBlackMax{5.62}

\PreprintCoverPaperTitle{Search for new phenomena in the dijet mass distribution
using $pp$ collision data at $\sqrt{s}=8$~TeV with the ATLAS detector}

\PreprintIdNumber{CERN-PH-EP-2014-147}

\PreprintCoverAbstract{Dijet events produced in LHC proton-proton collisions at a
center-of-mass energy $\sqrt{s}=8$~TeV are studied
with the ATLAS detector using the full 2012 data set, with an integrated
luminosity of \integLumi. Dijet masses up to about $\HighestDijetMass$~TeV
are probed. No resonancelike features are observed in the dijet mass spectrum.
Limits on the cross section times acceptance
are set at the 95\% credibility level for various hypotheses of new phenomena
in terms of mass or energy scale, as appropriate. This analysis excludes
excited quarks with a mass below \MjjLimitqstar~TeV,
color-octet scalars with a mass below \MjjLimits8~TeV,
heavy $\Wprime$ bosons with a mass below \MjjLimitWprime~TeV,
chiral $\Wstar$ bosons with a mass below \MjjLimitWstar~TeV,
and quantum black holes with six extra space-time dimensions
with threshold mass below \MjjLimitQBH~TeV.}

\PreprintJournalName{Phys. Rev. D}

\hypersetup{linkcolor=blue,citecolor=blue,filecolor=black,urlcolor=blue}

\makeatletter
\def\NAT@def@citea{\def\@citea{\NAT@separator}}
\makeatother

\makeatletter
\newcommand\footnoteref[1]{\protected@xdef\@thefnmark{\ref{#1}}\@footnotemark}
\makeatother

\begin{document}
\title{
Search for new phenomena in the dijet mass distribution
using $pp$ collision data at $\sqrt{s}=8$~TeV with the ATLAS detector
}

\author{ATLAS Collaboration}

\begin{abstract}
Dijet events produced in LHC proton-proton collisions at a
center-of-mass energy $\sqrt{s}=8$~TeV are studied
with the ATLAS detector using the full 2012 data set, with an integrated
luminosity of \integLumi. Dijet masses up to about $\HighestDijetMass$~TeV
are probed. No resonancelike features are observed in the dijet mass spectrum.
Limits on the cross section times acceptance
are set at the 95\% credibility level for various hypotheses of new phenomena
in terms of mass or energy scale, as appropriate. This analysis excludes
excited quarks with a mass below \MjjLimitqstar~TeV,
color-octet scalars with a mass below \MjjLimits8~TeV,
heavy $\Wprime$ bosons with a mass below \MjjLimitWprime~TeV,
chiral $\Wstar$ bosons with a mass below \MjjLimitWstar~TeV,
and quantum black holes with six extra space-time dimensions
with threshold mass below \MjjLimitQBH~TeV.
\end{abstract}
\maketitle

\section{Introduction}

This paper describes the search for new phenomena (NP) in the two-jet (dijet) invariant mass spectrum
in the full 2012 data set delivered by the Large Hadron Collider (LHC) at CERN, and collected
with the ATLAS detector.
The studies reported here select events containing two or more jets.
The two highest-\pT{} (``leading'' and ``subleading'') jets are combined
to determine the dijet invariant mass, $\mjj$.
High-transverse-momentum (high-\pT{}) dijet events are produced copiously by QCD processes, and can reach the
highest mass scales accessible in LHC $pp$ collisions.
The QCD processes, along with a subpercent admixture of additional Standard Model (SM)
processes, create a smooth rapidly falling spectrum in $\mjj$.
Many NP models describe new particles or excitations
created as $s$-channel resonances with appreciable branching
ratios to final states involving quarks and gluons ($q$ and $g$), and can produce dijet final states.
If the resonance width is sufficiently narrow, these NP signals would appear
as local excesses (bumps) in the dijet mass spectrum over the smooth SM background.

Studies searching for excesses in dijet mass spectra have been performed in all
previous collider experiments, including CDF and D0 at the
Tevatron\cite{Arnison1986244,Bagnaia1984283,CDF:2009DijetSearch,DZero:2009DijetAng}.
At the LHC, the CMS and ATLAS experiments have continued this program,
starting from the first 2010 data
\cite{ATLAS:2010bc,ATLAS:2010eza,CMS:2010dijetmass,CMS:2010centrality,
      CMS:2011DijetMassAngle,CMS:Res2011,Aad:2011aj,ATLAS:Res2011,
      ATLAS:2012pu,Chatrchyan:2013qha}.
The most recent published ATLAS results used the 2011 data set of 4.8 \invfb{}
at a center-of-mass energy of $\sqrt{s}=7$~TeV~\cite{ATLAS:2012pu}.
In 2012, the LHC delivered an integrated
luminosity of \integLumi\cite{Aad:2013ucp}
in proton-proton ($pp$) collisions,
roughly a factor of 4 more than used for previous dijet studies.
In addition, the increase in center-of-mass energy
to $\sqrt{s}=8$~TeV in 2012 increased the sensitivity in searches for new
phenomena, and pushed exclusion limits to higher masses and energy scales.
This increased kinematic reach, combined with a new online event selection strategy
employed in the present analysis, provides the largest dijet invariant mass range coverage
to date, from 250 GeV to 4.5 TeV.

No significant excess is observed above the background.
A number of NP models are compared to data to derive limits.
The models chosen span a range of characteristic masses (or energy
scales) and cross sections, and are complementary in terms of the flavor of their final-state partons.
The benchmark models under consideration include excited quarks ($\qstar$) decaying to
$qg$~\cite{Baur:1987ga,Baur:1989kv}, color-octet scalars ($s8$) decaying to
$gg$~\cite{Frampton1987a,Frampton1987b,Bagger1988,Han:2010rf},
heavy $\Wprime$ gauge bosons decaying to $q\bar{q}'$~\cite{Georgi1990541,Grojean:2011vu,RHW1,RHW2,Wprime1,Wprime2,Wprime3,Wprime4},
two forms of chiral $\Wstar$ gauge bosons~\cite{Wstar3,Wstar2,Wstar1,WstarOptim} and
quantum black holes~\cite{Feng:2004,Meade:2007sz,Calmet:2008dg,Gingrich:2009hj}
decaying to a mixture of quarks and gluons.
The current dijet search also sets limits on new generic resonances whose intrinsic width is convolved with effects due to parton distribution functions (PDFs), parton shower, nonperturbative effects and detector resolution.

\section{The ATLAS detector}
\label{sec:atlasDetector}

A detailed description of the ATLAS detector has been published
elsewhere~\cite{DetectorPaper}.  The detector is instrumented over almost
the entire solid angle around the $pp$\ collision point with layers of
tracking detectors, calorimeters, and muon chambers.

In ATLAS, high-\pT{} hadronic jets are measured using
a finely segmented calorimeter system, designed to achieve
high reconstruction efficiency and excellent energy resolution.
Electromagnetic calorimetry is provided by high-granularity
liquid-argon (LAr) sampling calorimeters, using lead as an absorber.
The calorimeters are split into a barrel region
($|\eta|<1.475$)
and two end-cap ($1.375<|\eta|<3.2$) regions.\footnote{ATLAS uses a right-handed coordinate system with its origin at the nominal interaction point (IP) in the centre of the detector and the $z$-axis along the beam pipe. The $x$-axis points from the IP to the centre of the LHC ring, and the $y$-axis points upward. Cylindrical coordinates $(r,\phi)$ are used in the transverse plane, $\phi$ being the azimuthal angle around the beam pipe. The pseudorapidity is defined in terms of the polar angle $\theta$ as $\eta=-\ln\tan(\theta/2)$.}

The hadronic calorimeter is divided into barrel, extended barrel ($|\eta|<1.7$)
and end-cap ($1.5<|\eta|<3.2$) regions. The barrel and extended
barrel are instrumented with scintillator tiles and steel absorbers,
while the end-caps use copper with LAr modules.
The forward calorimeter ($3.1<|\eta|<4.9$)
is instrumented with modules using LAr as the active medium and copper or tungsten as absorbers
to provide electromagnetic and hadronic energy measurements, respectively.

\section{Jet and event selection}
\label{sec:jetCalibration}

Jets are reconstructed from contiguous groups of calorimeter cells that
contain significant energy above noise
(topological clusters)~\cite{ATLAS-LARG-PUB-2008-002}.
The anti-$k_t$ jet algorithm~\cite{antikT,Cacciari:2006} is applied to
these clusters using a distance parameter of $R = 0.6$.
Effects from additional $pp$ interactions in the same and neighboring
bunch crossings are corrected for using the calibration procedure described in Ref.~\cite{ATLAS-CONF-2013-083}.
Simulated QCD multijet events are used for the derivation of the jet calibration to the hadronic scale.
They are produced with the event generator \Pythia{}~\cite{Pythia8}~8.160, using the CT10 PDF~\cite{Lai:2010vv}
and the AU2 set of parameters to describe the nonperturbative effects tuned to ATLAS data (the AU2 tune)
~\cite{ATL-PHYS-PUB-2012-003}.
Detector effects are simulated using \Geant{} within
the ATLAS software infrastructure~\cite{Agostinelli:2002hh,ATLSIM}.
Additional simulated minimum-bias events are overlaid onto the hard scattering,
both within the same bunch crossings and within trains of consecutive bunches.
The same software used to reconstruct data is used for the Monte Carlo (MC) samples.
The level of agreement between data and MC simulation is further improved by the
application of calibration constants obtained with {\it in situ} techniques based on momentum balancing between central and forward jets, between photons or $Z$ bosons and jets, and between high-momentum jets and a recoil system of low-momentum jets~\cite{ATLAS-CONF-2013-004}.

The energy scale of central jets relevant to this search
is known to within 4\%~\cite{ATLAS-CONF-2013-004}.
The jet energy resolution is estimated both in data and in simulation using transverse
momentum balance studies in dijet events~\cite{Aad:2012ag}, and are found to
be in agreement within uncertainties.
The dijet mass resolution is approximately
\MjjResolutionLow{} at 200 GeV, and improves to less than
\MjjResolutionHigh{} above 2 TeV.
The measured dijet mass distribution is not corrected for detector
resolution effects.

The data were collected using single-jet
triggers~\cite{Aad:2012xs}.
These triggers are designed to select events that have at least
one large transverse energy deposition in the calorimeter.
To match the data rate to the processing and storage capacity available
to ATLAS, the triggers with low-\pT\ thresholds are prescaled;
only a preselected fraction of all events passing
the threshold is recorded.
Combinations of prescaled single-jet triggers are used to reach lower
dijet invariant masses. The highest prescale for the trigger combinations
used in this analysis is 1/460000. This trigger combination is used
for jets with \pT\ between 59 and 99 GeV.

For a given leading-jet \pT{},
a predetermined combination of triggers with efficiencies exceeding 99.5\%
is used to select the event. Each event is weighted according to the
average effective integrated luminosity recorded by the given
trigger combination~\cite{CombTriggers}.

\begin{figure}[!htb]
\centering
\includegraphics[width=0.45\textwidth]{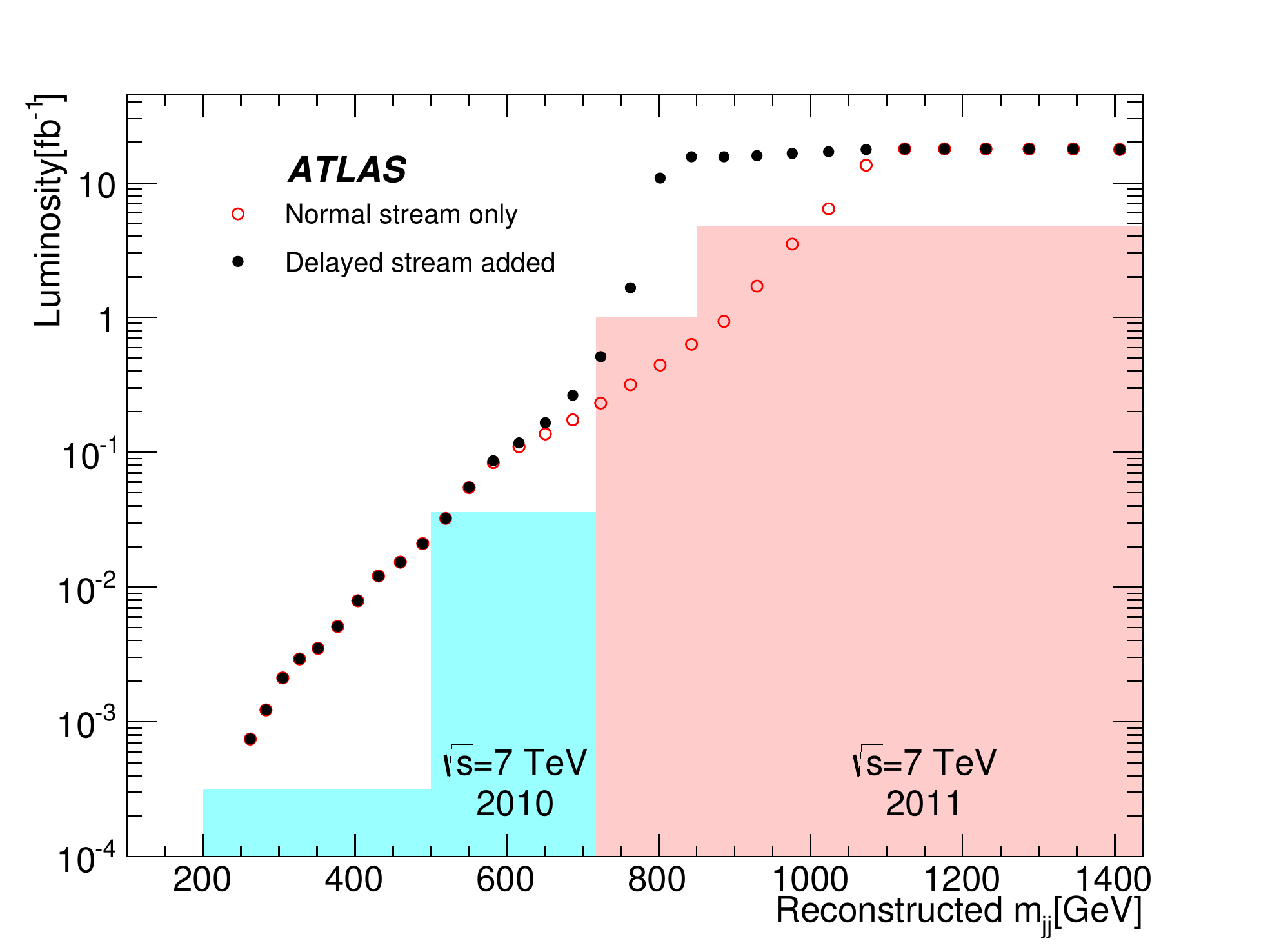}
\caption{Recorded effective integrated luminosity as a function of dijet
         mass for all former ATLAS dijet searches (shaded boxes).
         The integrated luminosity per dijet mass bin from the 2012 data used in the current analysis
         is shown without (open circles) and with (filled circles)
         the added delayed data stream.
}
\label{fig:DelayedStreamEffLumi}
\end{figure}

During the 2012 data taking, ATLAS recorded data at a rate that was higher than
the rate of the offline reconstruction:
400 Hz of recorded data were promptly reconstructed, while 100 Hz of data from hadronic triggers were recorded and reconstructed later (the ``delayed stream'').
Dijet events from the delayed data stream fall primarily into the region between 750\,GeV and 1\,TeV.
They are used to further increase the size of the analysis data set as follows.
First, two independent data sets are built from the delayed and normal trigger streams.
The $\mjj$ distributions from the two data sets have checked to be in agreement with a shape-only Kolmogorov-Smirnov test,
leading to a probability of 86\% for their compatibility.
The two dijet mass distributions measured from these data sets are then
averaged, using the effective integrated luminosity as a weight.
The delayed stream increases the luminosity recorded in this region of phase space
by up to an order of magnitude.
The effective integrated luminosity attained in this analysis
(using the 2012 normal stream with the added delayed stream)
is compared to previous ATLAS analyses in Fig.~\ref{fig:DelayedStreamEffLumi}.

Events used for the search are required to have at least one collision
vertex defined by two or more charged-particle tracks. In the presence of
multiple $pp$ interactions, the collision vertex with the largest scalar
sum of $p_{\mathrm{T}}^2$ of its associated tracks is chosen as the primary vertex.

Events are rejected if either the leading jets, or any of the other jets with
\pT\ greater than 30\% of the \pT\ of the subleading jet, are poorly
measured or have a topology characteristic of noncollision background or
calorimeter noise~\cite{JetEtMiss_Cleaning_Note}.
Poorly measured jets correspond to energy depositions in regions where the energy
measurement is known to be inaccurate. Events are also rejected if
one of the jets relevant to this analysis falls into regions of the
calorimeter that were nonoperational during data taking. An inefficiency of roughly 10\%
due to this veto is emulated in MC signal samples following the same conditions as data.
Central values and statistical errors of the dijet mass spectra of both the data and MC
signal samples are scaled, in order to correct for this inefficiency.

Additional kinematic selection criteria are used to enrich the dijet sample
with events in the hard-scatter region of phase space.
The rapidity $y$ of the two leading jets must be within $|y| < 2.8$.
The leading and subleading jets are required to have a \pT{} $> 50$ GeV,
ensuring a jet reconstruction efficiency of 100\%~\cite{Romeo:1334193}
both for QCD background and for all benchmark models under consideration.
Events must satisfy $|y^*| = \frac{1}{2}|y_{\text{lead}} - y_{\text{sublead}}| < 0.6$\ and $\mjj > 250$~GeV.
The invariant mass cut of $\mjj > 250$~GeV is chosen such that the dijet mass spectrum
is unbiased by the kinematic selection on \pT{}.

\section {Comparison of the dijet mass spectrum to a smooth background}
\label{sec:smoothfit}

\begin{figure}[!htb]
  \centering \includegraphics[width=0.45\textwidth]{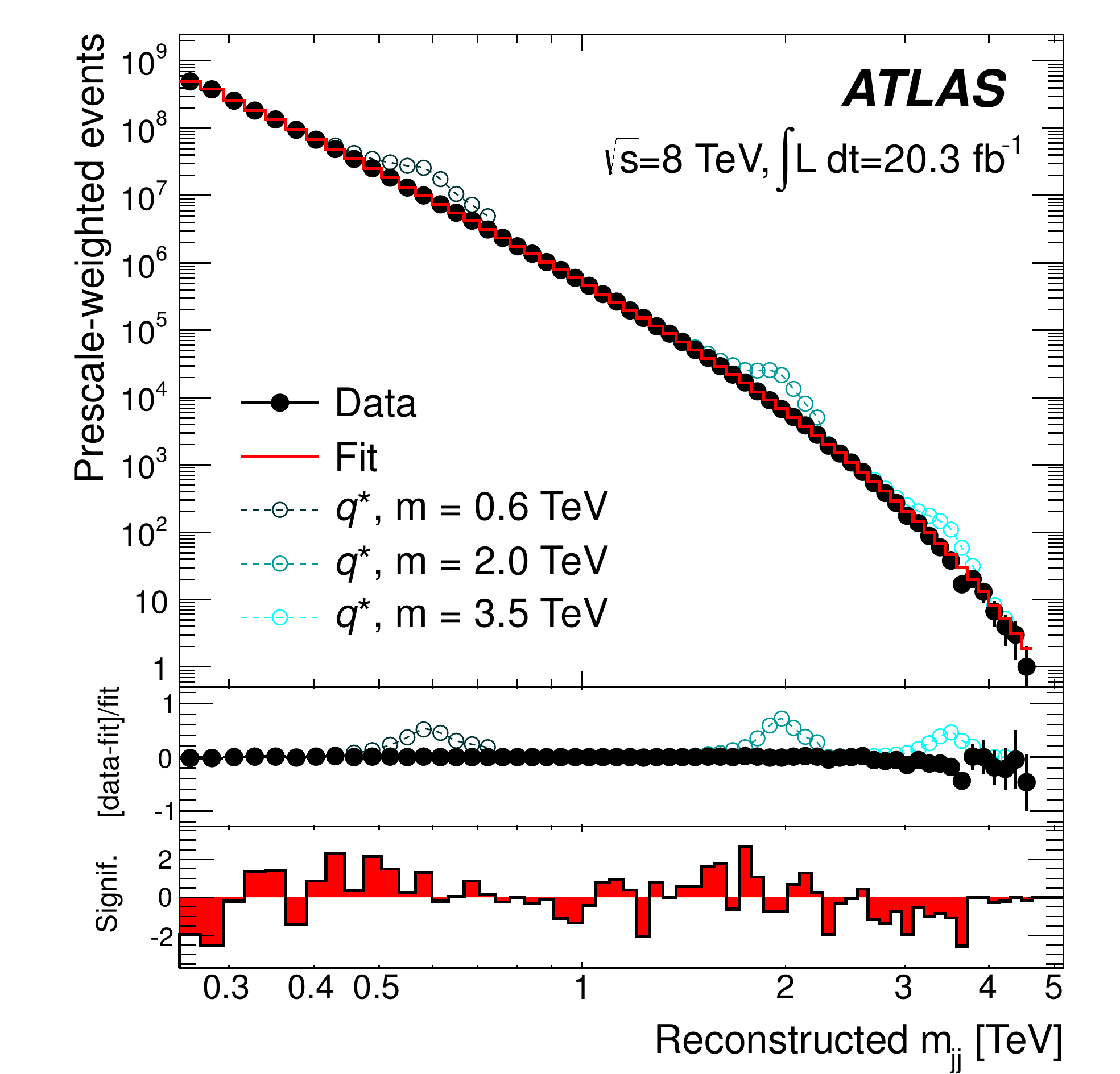}
  \caption{The reconstructed dijet mass distribution (filled points) fitted with
  a smooth functional form (solid line).
  Predictions for three $\qstar$ masses are shown above the background.
  The central panel shows the relative difference between the data and the background fit
  with overlaid predictions for the same $\qstar$ masses.
  The bin-by-bin significance of the data-background difference considering statistical uncertainties only is shown in the bottom panel.}
  \label{fig:massdist}
\end{figure}

The observed dijet mass distribution in data, after all selection requirements, is shown
in Fig.~\ref{fig:massdist}. The bin width varies with mass and
is chosen to approximately equal the dijet mass resolution
derived from simulation of QCD processes.
The predictions for an excited quark $\qstar$ with
three different mass hypotheses are also shown.

The search for resonances in $\mjj$\
uses a data-driven background estimate derived by fitting a smooth
functional form to the spectrum. An important feature of this functional
form is that it allows for smooth background variations, but
does not accommodate localized excesses that could indicate the
presence of NP signals.
In previous studies, ATLAS and other experiments
~\cite{doi:10.1142/S0217751X11054905}
have found that the following function provides a satisfactory fit to the QCD
prediction of dijet production:

\begin{equation}
  f(x) = p_1 (1 - x)^{p_2} x^{p_3 + p_4\ln x},
\label{Eq:fitfunction}
\end{equation}

where the $p_i$\ are fit parameters, and $x \equiv \mjj/\sqrt{s}$.
The uncertainty associated with the stability of the fit is
carried forward as a nuisance parameter in the statistical analysis.

The functional form was selected using a data set consisting of a quarter of the full data,
a quantity known to be insensitive to resonant new physics at dijet masses above 1.5 TeV
after the previous public result on 13 \invfb\ of data~\cite{ATLAS-CONF-2012-148}.
A range of parametrizations were tested on the blinded data set using a k-fold cross-validation and
there was found to be no substantial difference between the standard function of Eq.~\ref{Eq:fitfunction}
and higher-order parametrizations, so the function with a simpler form and a published precedent was selected.
The $\chi^2$-value of the fit to the blinded data set was \BlindedFitChiSquare{} for \FitNDF{} degrees of freedom using the
parameterisation of Eq.~\ref{Eq:fitfunction}.
The fit function showed good agreement to both the fully simulated dijet mass spectrum obtained
from the simulated \Pythia{} 8.160 QCD multijet events mentioned in Sec.~\ref{sec:jetCalibration}, corrected for
next-to-leading-order effects using the \NLOJET\ v4.1.3 program~\cite{Nagy1,catani-1998-510,*Catani1998503}
as described in Ref.~\cite{Aad:2011aj}, and from a large-statistics sample of generator-level events,
for which the $chi^2$ of the fit was 58 for 55 degrees of freedom.
While the number of data events is matched or surpassed by the number of fully simulated events
starting from dijet masses of roughly 2 TeV, the generator-level statistics is sufficient to reproduce that of data.
The $\chi^2$-value of the fit to data shown in Fig.~\ref{fig:massdist}
is \FitChiSquare{} for \FitNDF{} degrees of freedom.

\interfootnotelinepenalty=1000

The center panel of Fig.~\ref{fig:massdist}
shows the relative difference between the data and the background fit, and overlays the shapes
that would be expected in the presence of three sample $\qstar$ signals.
The bottom panel of Fig.~\ref{fig:massdist} shows the significance
of the difference between the data and the fit
in each bin. The significance is calculated taking only statistical uncertainties
into account, and assuming that the data follow a Poisson distribution with the
expected value given by the fit function.

For each bin a $p$-value is determined by assessing the probability of a background fluctuation
leading to a number of events higher than or equal to the observed excess, or lower than or equal
to the observed deficit.
This $p$-value is converted to a significance in terms
of an equivalent number of standard deviations (the $z$-value)~\cite{PlotDiffs}.
Where there is an excess (deficit) in data in a given bin, the significance
is plotted as positive (negative).\footnote{
In mass bins with small expected number of events, where the observed number
of events is similar to the expectation, the Poisson probability of a
fluctuation at least as high (low) as the observed excess (deficit) can be
greater than 50\%, as a result of the asymmetry of the Poisson distribution.
Since these bins have too few events for the significance to be meaningful,
these bins are drawn with zero content.}
To test the degree of consistency between the data and the fitted background,
the $p$-value of the fit is determined by calculating the $\chi^2$-value
from the data and comparing this result to the $\chi^2$ distribution obtained
from pseudoexperiments drawn from the background fit, as described in a
previous publication~\cite{Aad:2011aj}.  The resulting $p$-value is \ChiSqPvalue{}.

The \BumpHunter\ algorithm~\cite{Aaltonen:2008vt,Choudalakis:2011bh}
is used to establish the presence or absence of a narrow resonance in the dijet
mass spectrum, as described in greater detail in previous
publications~\cite{Aad:2011aj,ATLAS:Res2011}.  Starting with a two-bin window,
the algorithm increases the signal window and shifts its location until
all possible bin ranges, up to the widest window corresponding to
half the mass range spanned by the data, are tested.
The most significant excess of data above the smooth spectrum
(``bump'') is defined by the set of bins over which the integrated excess
of data over the fit prediction has the smallest probability
of arising from a background fluctuation, assuming Poisson statistics.

The \BumpHunter\ algorithm accounts for the so-called ``look-elsewhere effect''
\cite{lookelsefirstcite}, by performing a series of pseudoexperiments
drawn from the background estimate to determine the probability that random
fluctuations in the background-only hypothesis would create an excess
anywhere in the spectrum at least as significant as the one observed.
Furthermore, to prevent any NP signal from biasing the background estimate,
if the most significant local excess from the background fit has a $p$-value
smaller than 0.01, the corresponding region is excluded and a new background fit is
performed. The exclusion is then progressively widened bin by bin until
the $p$-value of the remaining fitted region is acceptable.
No such exclusion is needed for the current data set.

The most significant discrepancy identified by the \BumpHunter\ algorithm
in the observed dijet mass distribution in Fig.~\ref{fig:massdist}
is a seven-bin excess in the interval \LowEdgeHighestBump{}--\HighEdgeHighestBump{}~GeV.
The probability of observing an excess at least as large somewhere in the mass spectrum for
a background-only hypothesis is~\BHPvalue{}, corresponding to a $z$-value of~\BHZvalue{}$\sigma$.
To conclude, this test shows no evidence for a resonant signal in the observed $\mjj$\ spectrum.

\section {Simulation of hypothetical new phenomena}
\label{sec:SimulNP}

In the absence of any significant signals indicating the presence of phenomena
beyond the SM, Bayesian 95\% credibility level (C.L.) limits are determined for a
number of NP hypotheses that would produce localized excesses.

Samples for NP models are produced by a variety of event generators. The partons
originating from the initial $2\to 2$ matrix elements are passed to \Pythia~8.160 with
the AU2 tune~\cite{ATL-PHYS-PUB-2012-003}. \Pythia\ uses~\pT{}-ordered parton
showers to model additional radiation in the leading-logarithm approximation~\cite{pythiapartonshower}.
Multiple parton interactions~\cite{Sjostrand:2004ef}, as well as fragmentation and hadronization
based on the Lund string model~\cite{lundstring}, are also simulated.
Renormalization and factorization scales for the NP models
are set to the mean \pT\ of the two leading jets.

Excited $u$- and $d$- quarks ($\qstar$), one possible manifestation of quark compositeness, are
simulated in all decay modes using the \Pythia~8.162 generator,
using the CT10 PDF.
Excited quarks are assumed to decay to quarks via gauge couplings set to unity,
leading to a $qg$ final state approximately 83\% of the time
(the remaining generated decays involve W/Z or $\gamma$ emission). The acceptance,
defined as the fraction of generated events passing all reconstruction steps and the analysis selection
described in Sec.~\ref{sec:jetCalibration} using jets reconstructed from stable particles~\footnote{A stable particle is
defined as one that has a lifetime
longer than 10 ps.} excluding muons and neutrinos, is approximately 58\%.
The largest reduction in acceptance arises from the rapidity selection criteria.

The color-octet scalar model describes the production of exotic colored resonances ($s8$).
{\sc MadGraph}~5 (v1.5.5)~\cite{Alwall:2011uj} with the MSTW2008LO PDF~\cite{Martin:2009iq}
is employed to generate parton-level events at leading-order approximation.
Parton showering and nonperturbative effects are simulated with \Pythia~8.170.
Color-octet scalars can decay into two gluons and can then have a broader mass distribution
and larger tails than resonances decaying to quarks. For resonances produced
by this model, the acceptance ranges from 61\% to 63\%.

The production of heavy charged gauge bosons, $\Wprime$, has been sought through
decays to $q\bar{q}'$.  The specific model (sequential Standard Model, or SSM~\cite{Wprime1,Wprime2}) used in this
study assumes that the $\Wprime$ has $V-A$ SM couplings but
does not include interference between the $\Wprime$ and the $W$, leading
to a branching ratio to dijets of 75\%. The $\Wprime$
signal sample is simulated with the \Pythia~8.165 event generator using
the MSTW2008LO PDF. Instead of the LO
cross section values, the next-to-next-to-leading-order cross section values
calculated with the MSTW2008NNLO PDF are used in this analysis, as
detailed in~\cite{ATLAS:2014wra} and references therein.
The acceptance for $\Wprime$ bosons decaying to quarks ranges from 48\%
at masses below 1200 GeV to 40\% at 3200 GeV,
driven by the rapidity selection criteria.
At high $\Wprime$ masses, the acceptance decreases due to PDF suppression effects
causing the reconstructed dijet invariant mass to fall below the 250 GeV cut.

A new excited $\Wstar$ boson \cite{Wstar1, Wstar2}
is generated through a simplified model \cite{Wstar3} in the CalcHEP 3.4.2 generator,
in combination with the MSTW2008LO PDF and \Pythia~8.165 for
the simulation of nonperturbative effects.
The sine of the mixing angle in this model (sin$\phi_{X}$) is set to zero, producing leptophobic
decays of the $\Wstar$ that are limited to quarks. With sin$\phi_{X}$ = 1,
a leptophilic $\Wstar$ would instead be produced with branching ratios divided equally
between quarks (3 families $\times$ 3 flavors $\times$ 8.3\% BR = 75\%) and leptons (3 families $\times$ 8.3\% BR = 25\%).
The angular distribution of the $\Wstar$ differs from that of the other signals under study,
preferring decays with a wider separation in $y$.
The acceptance for both leptophobic and leptophilic $\Wstar$ spans 25\% to 27\%.

A model for quantum black holes (QBH) that decay to two jets is simulated using both the
\BlackMax~\cite{Dai:2007bm}\ and the \QBH\ \cite{Gingrich:2009da} generators,
to produce a simple two-body final-state scenario of quantum gravitational effects at the
fundamental Planck scale $M_{\rm D}$, with $n$ = 6 extra spatial dimensions in the context of the ADD model \cite{ArkaniHamed:1998nn}.
In this model, the Planck scale is set equal to the threshold mass for the quantum black hole production $m_{th}$.
These QBH models are used as benchmarks to represent any quantum gravitational
effect that produces events containing dijets.
The PDF used for the generation and parton shower of \BlackMax\ is CT10, while
the \QBH\ samples employ the MSTW2008LO PDF.
In the mass range considered, the branching ratio of QBH to dijets is above 85\% and
the acceptance is between 52\% and 55\% for \BlackMax\ and 54\%--56\% for \QBH{}.
nonperturbative effects for events coming from both event generators are simulated with \Pythia~8.170.

Further information on cross sections, branching ratio to dijets and acceptances for the
benchmark models under consideration can be found in HepData~\cite{HepData}.

All MC signal samples except for the excited quark signals are passed through a
fast detector simulation \cite{ATLFAST2-2011} with the jet calibration appropriately corrected to full simulation.
The excited quark signals are simulated using \Geant{} within the
ATLAS simulation infrastructure~\cite{ATLSIM}.

\section{Limits on new resonant phenomena from the $\mjj$ distribution}
\label{sec:mjj:reson}

The Bayesian method used for the limit setting is documented in Ref.~\cite{Aad:2011aj} and
implemented using the Bayesian Analysis Toolkit~\cite{2009CoPhC.180.2197C}.
Limits on the cross section times acceptance,
$\sigma\times {\cal A}$,
are set at the 95\% C.L. for the NP signal as a function of
$m_{\textrm{NP}}$, using a prior constant in signal strength and Gaussian priors for
the nuisance parameters due to the systematic uncertainties under consideration.
The full template shape is considered in the limit-setting procedure,
both in the fits to the data performed during the marginalization procedure~\footnote{The NP signal distribution
is added to the binned data spectrum, and the parameters of the background function are extracted by
fitting the combined distribution to a five-parameter function, where the fifth parameter is proportional
to the signal strength.} and in the likelihood for the determination of the 95\% C.L. limit.
The limit on $\sigma\times {\cal A}$ from data is interpolated linearly on the $x$ axis and logarithmically on the $y$ axis
between the mass points to create a continuous curve in signal mass.
The exclusion limit on the mass (or energy scale) of the
given NP signal occurs at the value of the signal mass where the limit
on $\sigma\times {\cal A}$ from data is the same as the
theoretical value, which is derived by interpolation between
the generated mass values.
This form of analysis is applicable to all resonant phenomena
where the NP couplings are strong compared to the scale of perturbative QCD
at the signal mass, so that interference between these terms can be neglected.

As in previous dijet resonance analyses, limits on dijet resonance production
are also determined using
a collection of hypothetical signal shapes that are assumed to
be Gaussian-distributed in $\mjj$. Signal shape templates are generated with means ($m_{\text{G}}$) ranging from
200~GeV to 4.0~TeV and with standard deviations ($\sigma_{\text{G}}$) corresponding
to the dijet mass resolution estimated from MC simulation and ranging from 7\% to 15\% of the mean.\footnote{
Limits are determined only for those Gaussian resonances whose means fall more than
$2\sigma$ from either edge of the data to preserve the stability of the
background estimation, so the limits from wider signals include
fewer mass points near the ends of the range.}
For further information on the mass resolution, see Appendix~\ref{sub:dijetmassresolution}.

An additional set of limits with minimal model assumptions is added to this publication.
For particles with a nonzero natural width generated at masses close to the collision energy,
the parton luminosity favors lower-mass collisions. This creates an asymmetric resonance
not well represented by a Gaussian distribution.
To handle this scenario, Breit-Wigner signals of fixed intrinsic widths
(0.5\% to 5\% of the resonance mass) are generated and multiplied by the parton
luminosities for different initial states ($qq$, $qg$, $gg$ and $q \bar{q}$) according
to the CT10 PDF. Effects of parton shower and nonperturbative effects are estimated using
\Herwig 2.6.3~\cite{Corcella:2000bw,Marchesini:1991ch}
and convoluted with the signal shape.

The detector resolution is accounted for by convolving the signal shape with
a Gaussian function of width equal to the detector resolution at each signal mass.
The result is then truncated below 250 GeV due to the dijet mass cut.
This produces a signal template shape that is still generic but more likely
to match the forms visible in actual physical processes.
The effect on the shape of the signal template originating from the $y^*$ cut is not
simulated in the signal templates used for these limits, due
to the possible model dependence of the angular distribution of the considered NP process.\footnote{
If a flat distribution in the jet polar angle in the center-of-mass rest frame for the new particle
is assumed for the NP model, the acceptance
can be calculated analytically and it amounts to roughly 56\% for all considered dijet masses.}
Tests of the benchmark model-specific templates indicate that the combined effect of the $y$ and $y^*$ acceptance requirement
is constant within 20\% as a function of the dijet mass, with the largest discrepancies outside the mass peak.

\subsection{Systematic uncertainties in limit setting}
\label{sec:systUnc}

The effects of several systematic uncertainties are considered when setting
limits on new phenomena. These are incorporated into the Bayesian marginalization
limit setting procedure using Gaussian priors, with one nuisance parameter for each uncertainty. They are listed below.

(1) Choice of fitting function: a tenfold cross validation~\cite{Kohavi95astudy} using the full data set shows that the background is also well described when introducing an additional degree of freedom to Eq.~(\ref{Eq:fitfunction}),

\begin{equation}
  f(x) = p_1 (1 - x)^{p_2} x^{p_3 + p_4\ln x + p_{5}(\ln x)^2}.
\label{Eq:alternativefitfunction}
\end{equation}
\noindent
The $\chi^2$-value of this five-parameter function fit to data
is 45 for 57 degrees of freedom.
Since the two fit functions provide background estimates that differ beyond statistical uncertainties,
an additional uncertainty is introduced due to the choice of fitting function.
The difference between the two background estimates is treated as a one-sided nuisance parameter, with a Gaussian
prior centered at zero corresponding to the background estimate from Eq.~(\ref{Eq:fitfunction})
and truncated to one-$\sigma$ corresponding to the background estimate from Eq.~(\ref{Eq:alternativefitfunction}).
The marginalized posterior indicates a preference for the alternative function.

 (2) Background fit quality: the uncertainty on the background parametrization from the fit is estimated by refitting bin-by-bin Poisson fluctuations of the data, as described in Ref.~\cite{Aad:2013cva}.
The resulting uncertainty is calculated by refitting a large number of pseudoexperiments drawn from the data, and defining the fit error from the variation in fit results in each bin: $\pm1\sigma$ in the uncertainty corresponds to the central 68\% of pseudoexperiment fit values in the bin.

 (3) Jet energy scale: shifts to the jet energy due to the various jet energy scale (JES) uncertainty components are propagated separately through the analysis of the signal templates. Changes in both shape and acceptance due to the JES uncertainty in the simulated signal templates are considered in the limit setting. Combined, the JES uncertainty shifts the resonance mass peaks by less than 3\%: this is the JES shift used for Gaussian and Breit-Wigner limits.

 (4) Luminosity: a 2.8\% uncertainty~\cite{Aad:2013ucp} is applied to the overall normalization of the signal templates.

 (5) Theoretical uncertainties: the uncertainty on the signal acceptance for the model-dependent limits due to the choice of PDF
 is derived employing the PDF4LHC recommendation~\cite{Botje:2011sn} using the envelope of the error sets of the NNPDF 2.1~\cite{Ball:2011mu} and MSTW2008LO. Renormalization and factorization scale uncertainties on the signal acceptance are considered
 for the $\Wprime$ and $s8$ signals but found negligible.
Since the $\Wprime$ cross section estimation used in this analysis includes NNLO corrections,
the uncertainties on cross section due to variations of the renormalization and factorization
scales, the choice of PDF, and PDF+$\alpha_{\rm s}$ variations on the
theoretical cross section are considered as well.

The effect of the jet energy resolution uncertainty is found to be negligible.
Similarly, effects due to jet reconstruction efficiency and jet angular resolution
lead to negligible uncertainties.

\begin{figure}[!htb]
  \centering \includegraphics[width=0.45\textwidth]{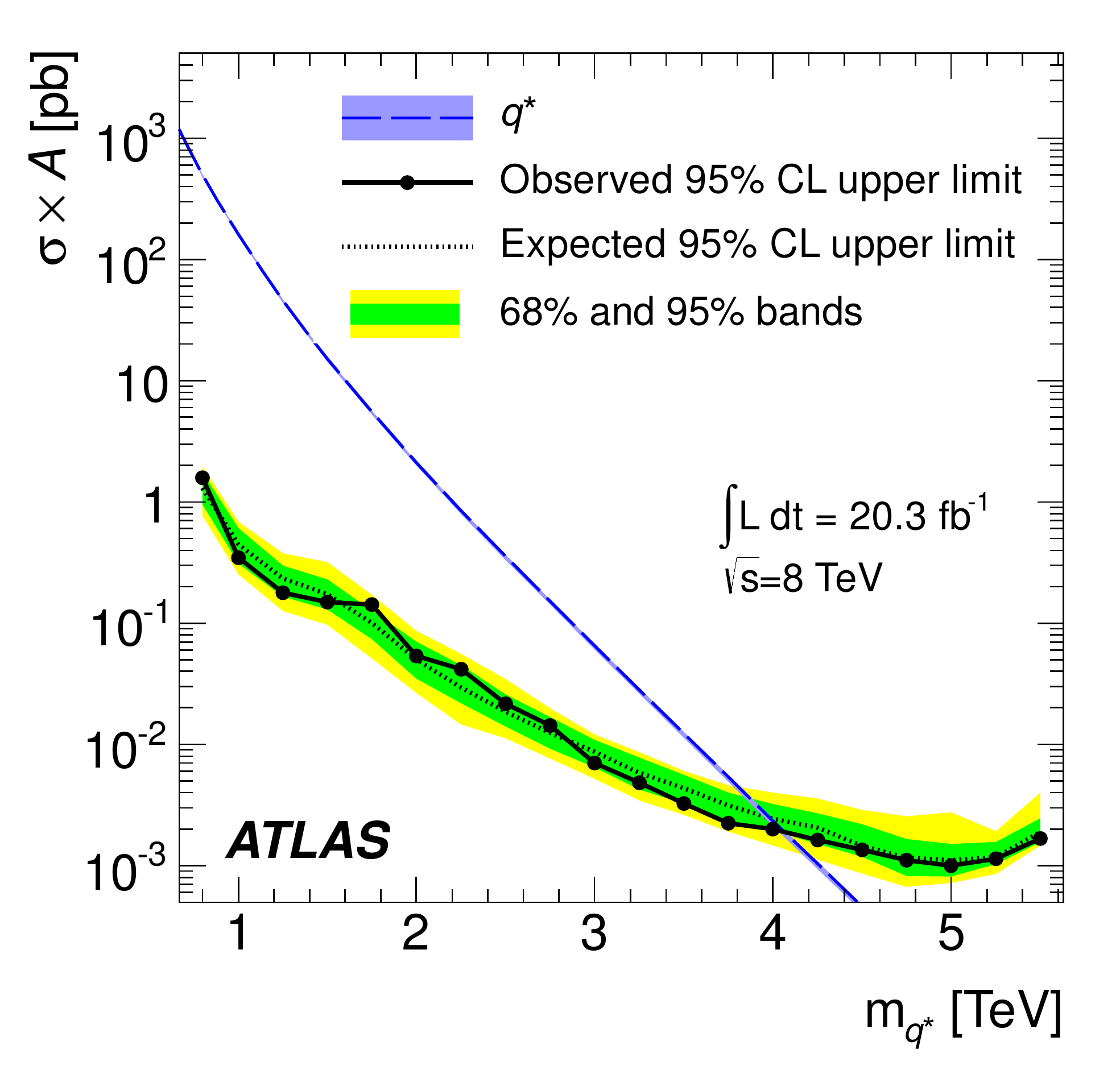}
  \caption{Observed (filled circles) and expected 95\% C.L. upper limits
      (dotted line) on $\sigma\times {\cal A}$
      for excited quarks as a function of particle mass.
	The green and yellow bands represent the
      68\% and 95\% contours of the expected limit.
      The dashed curve is the theoretical prediction of $\sigma\times {\cal A}$.
      The uncertainty on the nominal signal cross section due to the beam energy
      uncertainty is also displayed as a band around the theory prediction.
      The observed (expected) mass limit occurs at the crossing of the
      dashed $\sigma\times {\cal A}$ curve with the observed (expected)
      95\% C.L. upper limit curve.}
  \label{fig:limqstarmjj}
\end{figure}

\begin{figure}[!htb]
  \centering \includegraphics[width=0.45\textwidth]{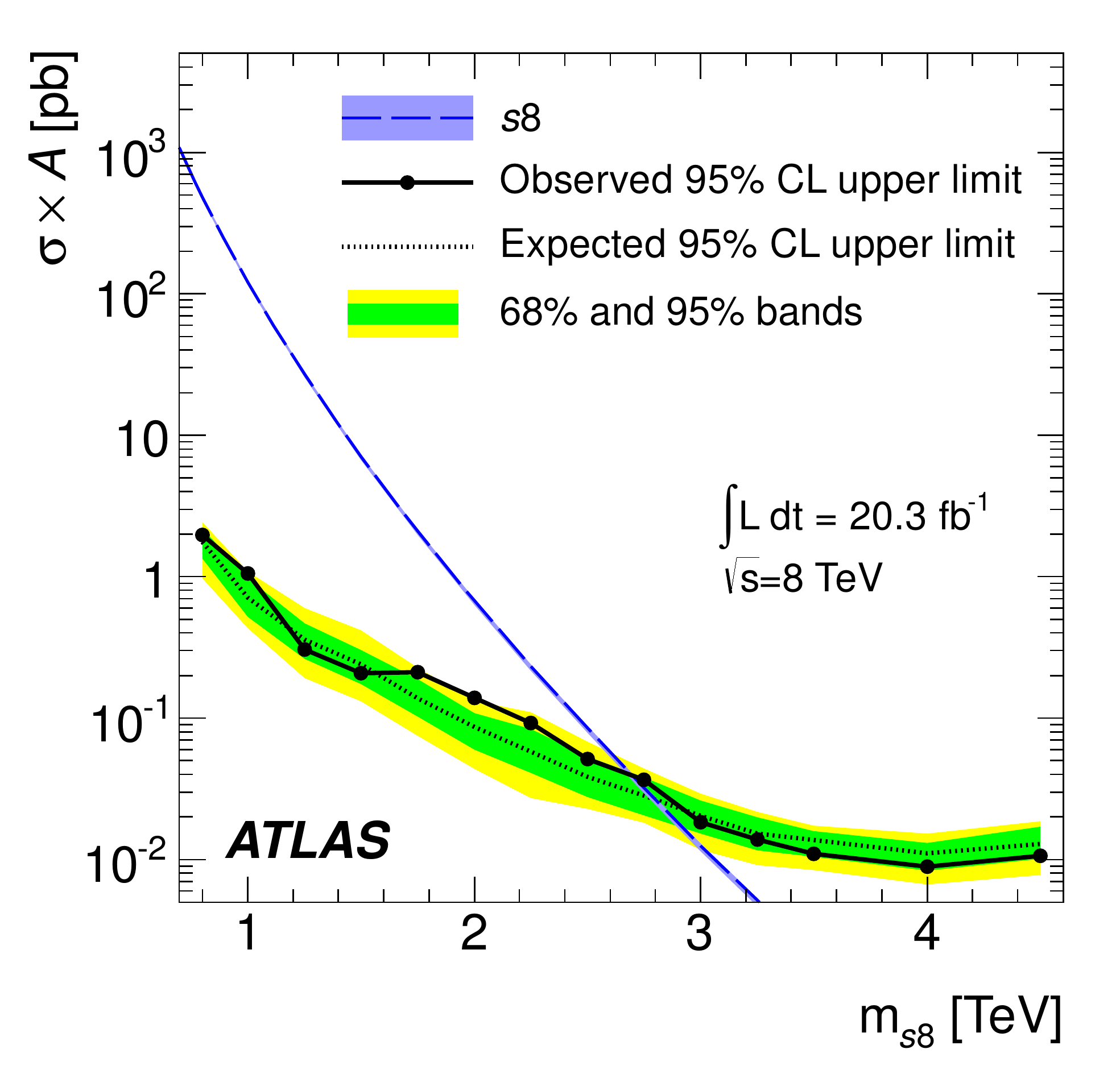}
  \caption{Observed (filled circles) and expected 95\% C.L. upper limits
      (dotted line) on $\sigma\times {\cal A}$
      for color-octet scalars as a function of particle mass.
	The green and yellow bands represent the
      68\% and 95\% contours of the expected limit.
      The dashed curve is the theoretical prediction of $\sigma\times {\cal A}$.
      The uncertainty on the nominal signal cross section due to the beam energy
      uncertainty is also displayed as a band around the theory prediction.
      The observed (expected) mass limit occurs at the crossing of the
      dashed $\sigma\times {\cal A}$ curve with the observed (expected)
      95\% C.L. upper limit curve.}
  \label{fig:lims8mjj}
\end{figure}

\begin{figure}[!htb]
  \centering \includegraphics[width=0.45\textwidth]{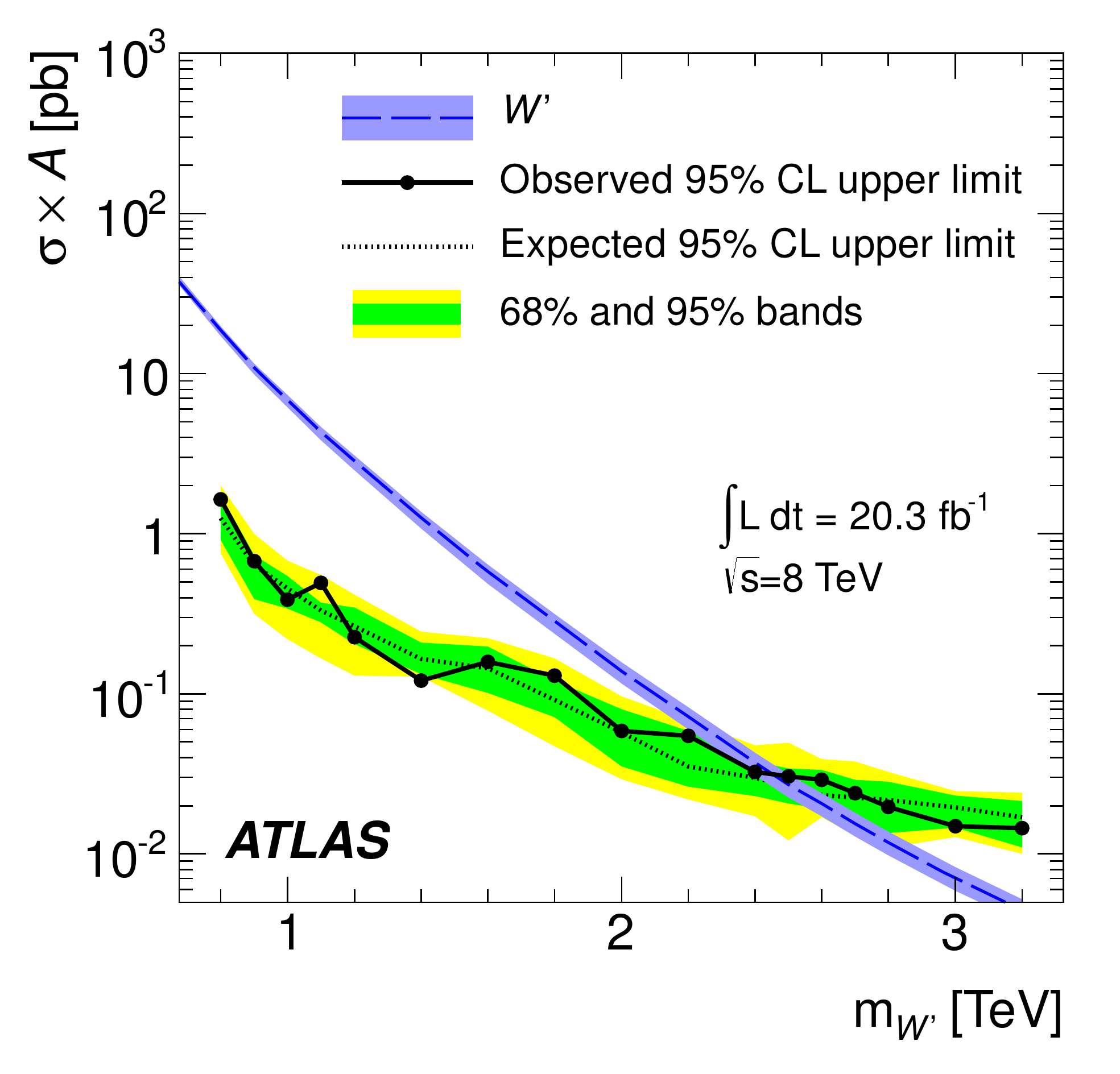}
  \caption{Observed (filled circles) and expected 95\% C.L. upper limits
      (dotted line) on $\sigma\times {\cal A}$
      for heavy vector bosons as a function of particle mass.
	The green and yellow bands represent the
      68\% and 95\% contours of the expected limit.
      The dashed curve is the theoretical prediction of $\sigma\times {\cal A}$.
      The uncertainty on the nominal signal cross section due to the beam energy
      uncertainty is also displayed as a band around the theory prediction.
      Additionally the uncertainty on the calculation of the next-to-next-to-leading
      order cross section is shown around the theory line.
      The observed (expected) mass limit occurs at the crossing of the
      dashed $\sigma\times {\cal A}$ curve with the observed (expected)
      95\% C.L. upper limit curve.}
  \label{fig:limwprimemjj}
\end{figure}

\begin{figure}[!htb]
  \centering \includegraphics[width=0.45\textwidth]{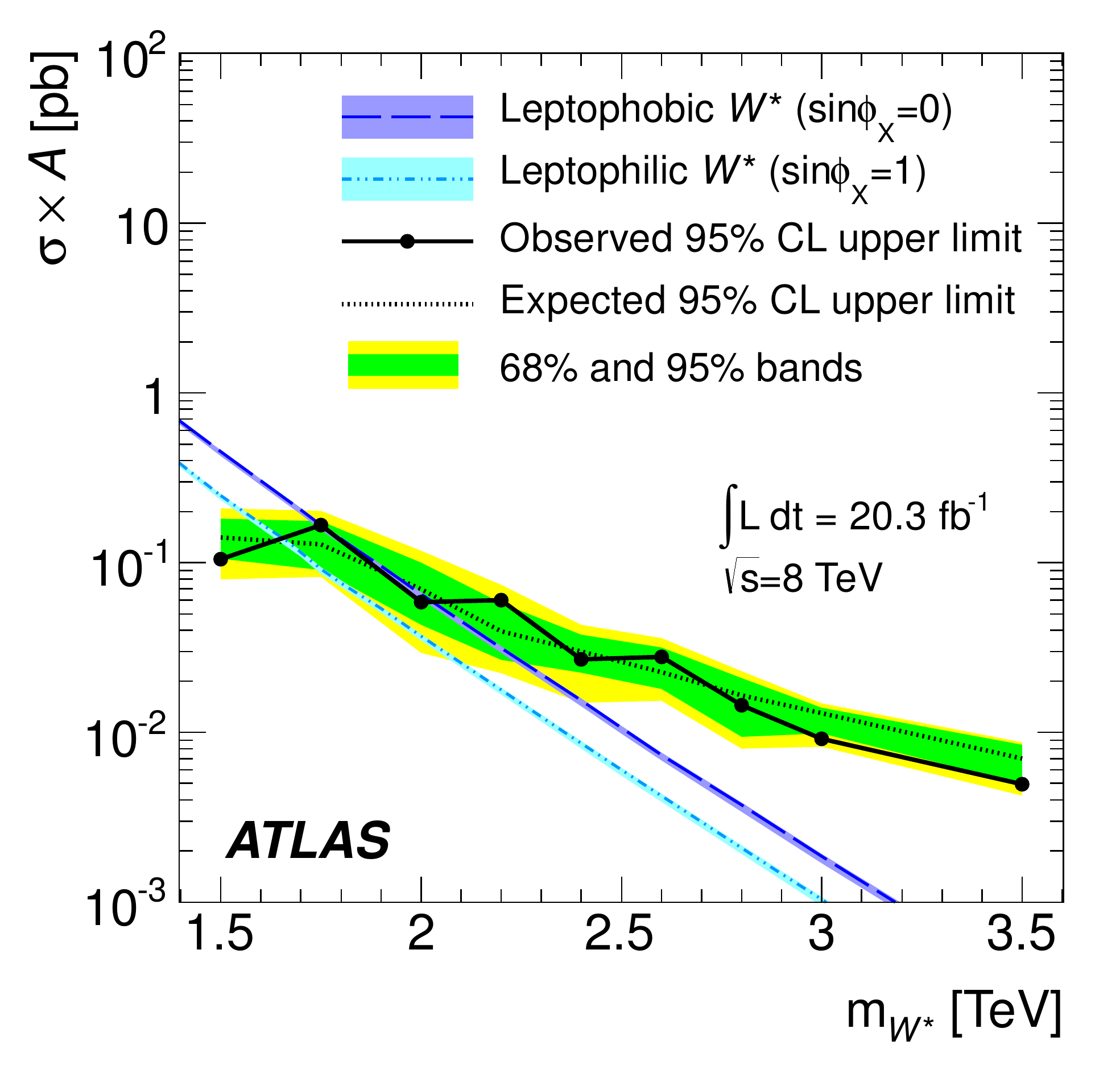}
  \caption{Observed (filled circles) and expected 95\% C.L. upper limits
      (dotted line) on $\sigma\times {\cal A}$
      for leptophobic and nonleptophobic excited vector bosons $\Wstar$ as a function of particle mass.
	The green and yellow bands represent the
      68\% and 95\% contours of the expected limit.
      The dashed curve is the theoretical prediction of $\sigma\times {\cal A}$.
      The uncertainty on the nominal signal cross section due to the beam energy
      uncertainty is also displayed as a band around the theory prediction.
      The observed (expected) mass limit occurs at the crossing of the
      dashed $\sigma\times {\cal A}$ curve with the observed (expected)
      95\% C.L. upper limit curve.}
  \label{fig:limwstarmjj}
\end{figure}

\begin{figure}[!htb]
  \centering \includegraphics[width=0.45\textwidth]{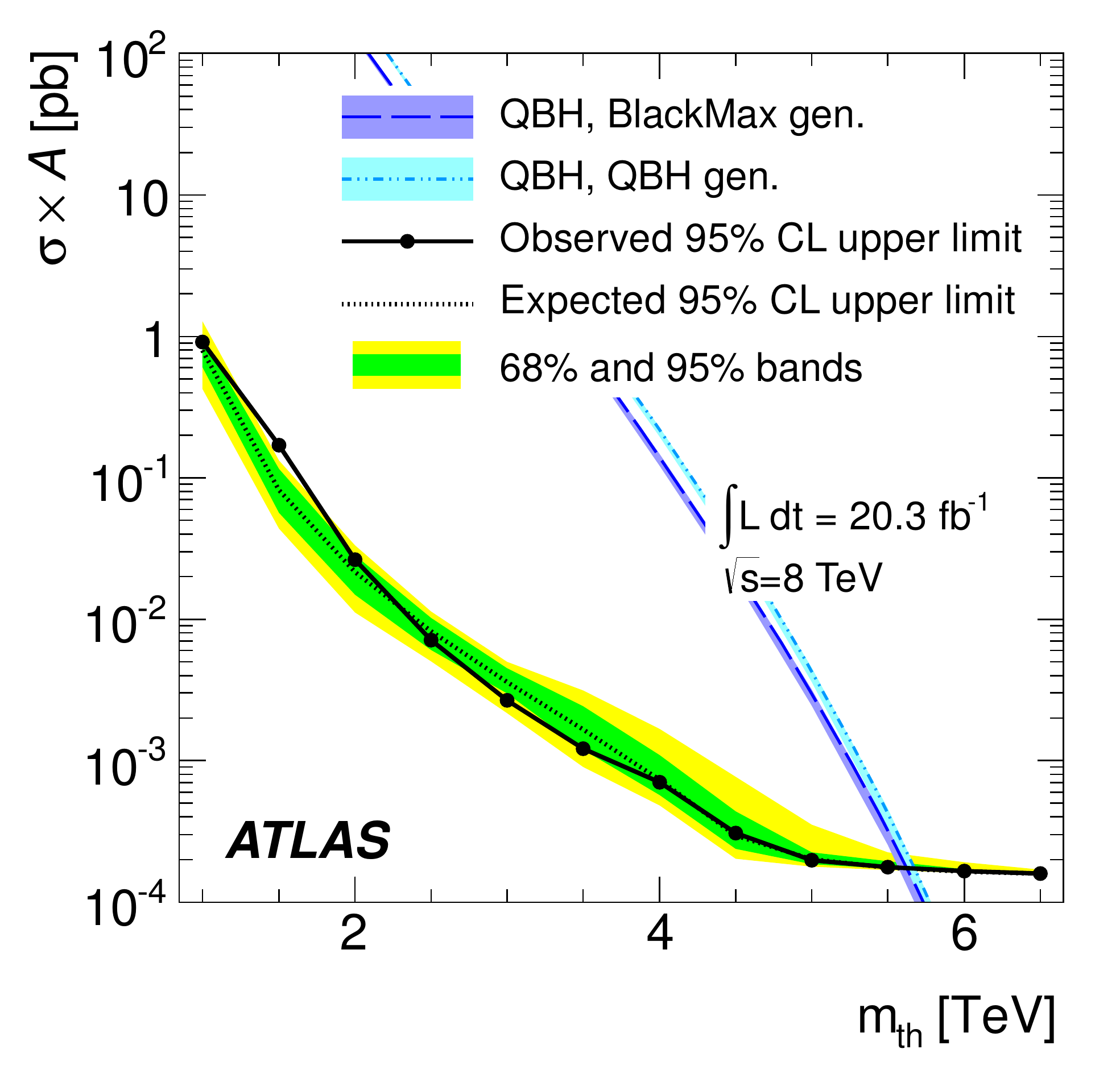}
  \caption{Observed (filled circles) and expected 95\% C.L. upper limits
      (dotted line) on $\sigma\times {\cal A}$
      for black holes simulated using the \textsc{Qbh} and \BlackMax\ generators as a function of particle mass.
	The green and yellow bands represent the
      68\% and 95\% contours of the expected limit.
      The dashed curve is the theoretical prediction of $\sigma\times {\cal A}$.
      The uncertainty on the nominal signal cross section due to the beam energy
      uncertainty is also displayed as a band around the theory prediction.
      The observed (expected) mass limit occurs at the crossing of the
      dashed $\sigma\times {\cal A}$ curve with the observed (expected)
      95\% C.L. upper limit curve.}
  \label{fig:limbmmjj}
\end{figure}

\subsection{Constraints on NP benchmark models}
\label{sec:modelDependent}

The resulting limits for excited quarks are shown in Fig.~\ref{fig:limqstarmjj}.
The expected lower mass limit at 95\% C.L. for $\qstar$ is \MjjLimitExpectedqstar~TeV,
and the observed limit is \MjjLimitqstar~TeV.
The limits for color-octet scalars are shown in Fig.~\ref{fig:lims8mjj}.
The expected mass limit at 95\% C.L. is \MjjLimitExpecteds8~TeV, and the observed limit
is \MjjLimits8~TeV.

The limits for heavy charged gauge bosons, $\Wprime$, are shown in Fig.~\ref{fig:limwprimemjj}.
The expected mass limit at 95\% C.L. is \MjjLimitExpectedWprime~TeV, and the observed limit
is \MjjLimitWprime~TeV.

The limits for the excited $\Wstar$ boson are shown in Fig.~\ref{fig:limwstarmjj}.
The plot shows the observed and expected limits calculated for a leptophobic $\Wstar$
but includes theory curves for both leptophobic and nonleptophobic $\Wstar$ given that
the acceptances for the two samples are the same to within 1\%.
The expected mass limit for the leptophobic model at 95\% C.L. is \MjjLimitExpectedWstar~TeV and
the observed limit is \MjjLimitWstar~TeV.
The expected mass limit for the nonleptophobic model at 95\% C.L. is \MjjLimitExpectedW1star~TeV and
the observed limit is \MjjLimitW1star~TeV.

The limits for black holes generated using \textsc{Qbh} and \textsc{BlackMax} are shown in
Fig.~\ref{fig:limbmmjj}. The observed limit is consistent between the two generators, but the
cross sections differ, hence the difference in the mass limit.  The observed limits for the two
models have visually matching shapes and normalizations, so only one (\textsc{BlackMax}) is selected for display.
The limits for both models are, however, computed separately and recorded.
The expected mass limit for \textsc{Qbh} black holes at 95\% C.L. is
\MjjLimitExpectedQBH~TeV, and the observed limit is \MjjLimitQBH~TeV.
For \textsc{BlackMax} black holes, the expected limit at 95\% C.L. is
\MjjLimitExpectedBlackMax~TeV and the observed limit is \MjjLimitBlackMax~TeV.
Above~$\sim$\HighestDijetMass~TeV the observed and expected limits are driven by the absence of any observed data events,
leading to identical observed and expected mass limits.

Although the search phase of the analysis starts at 250 GeV, $\sigma\times {\cal A}$ exclusion limits on benchmark NP models are set
starting at 800 GeV for the $\qstar$, $s8$, and $W'$ models, and at 1500 GeV for the $\Wstar$ model.
In the first three cases, this ensures that the rapid increase in the delayed stream statistics from 800 GeV
onwards does not shift the search to be more sensitive to the tails of the model, rather than to its peak.
In the $\Wstar$ model, the limited acceptance distorts the peak shape below 1500 GeV so that it cannot be
adequately treated as a resonance.
Exclusion limits on quantum black holes are set starting from 1 TeV
in light of the large cross section and of previous exclusion limits~\cite{Aad:2013cva, Aad:2013gma}.

\subsection{Generic resonance limits on dijet production}
\label{Sec:ModelIndep}

\begin{figure}[!htb]
  \centering
  \includegraphics[width=0.45\textwidth]{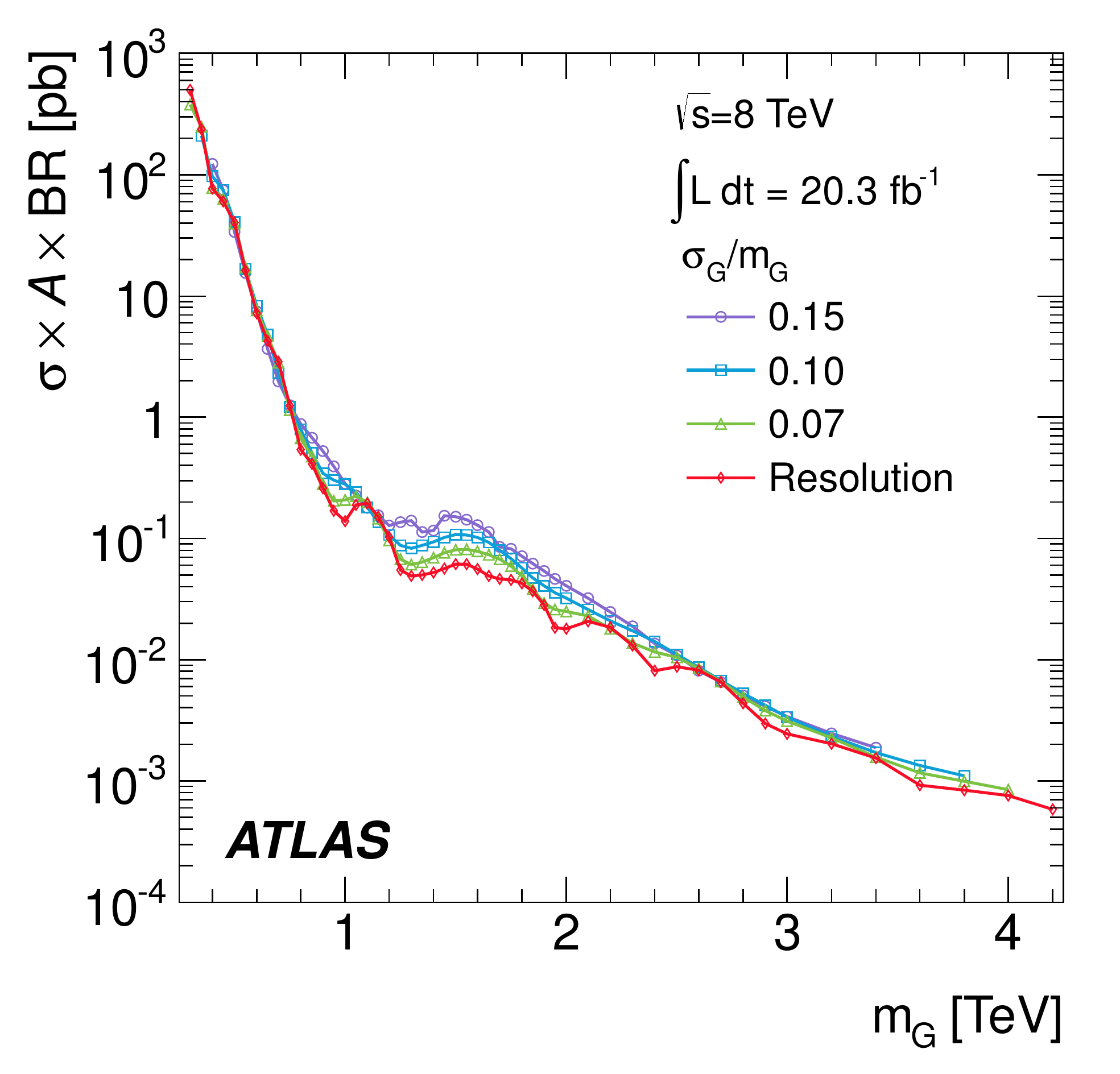}
  \caption{The 95\%\ C.L. upper limits on $\sigma\times {\cal A}$ for a simple Gaussian
           resonance decaying to dijets as a function of the mean mass, $m_{\text{G}}$, for
           four values of $\sigma_{\text{G}}/m_{\text{G}}$, taking into account
           both statistical and systematic uncertainties.}
  \label{fig:modindeplims}
\end{figure}

\begin{figure}[!htb]
  \centering
  \includegraphics[width=0.45\textwidth]{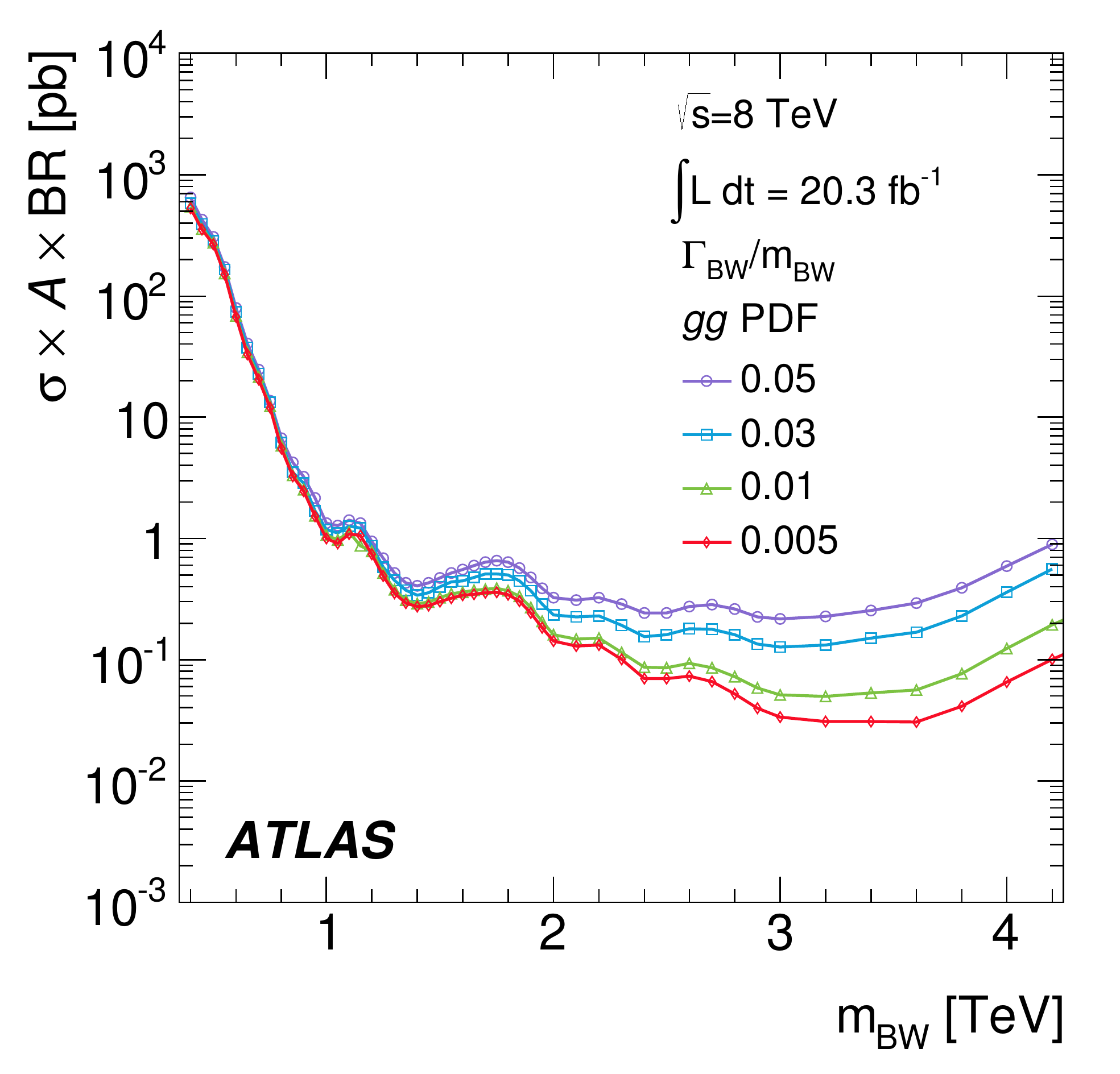}
  \caption{The 95\%\ C.L. upper limits on $\sigma\times {\cal A}$ for a Breit-Wigner narrow resonance
           produced by a $gg$ initial state
           decaying to dijets and convolved with PDF effects, dijet mass acceptance, parton shower and nonperturbative effects and detector resolution,
           as a function of the mean mass, $m_{\text{BW}}$, for different values of intrinsic width over mass ($\Gamma_{\text{BW}}/m_{\text{BW}}$),
           taking into account both statistical and systematic uncertainties.}
  \label{fig:BWGG}
\end{figure}

\begin{figure}[!htb]
  \centering
  \includegraphics[width=0.45\textwidth]{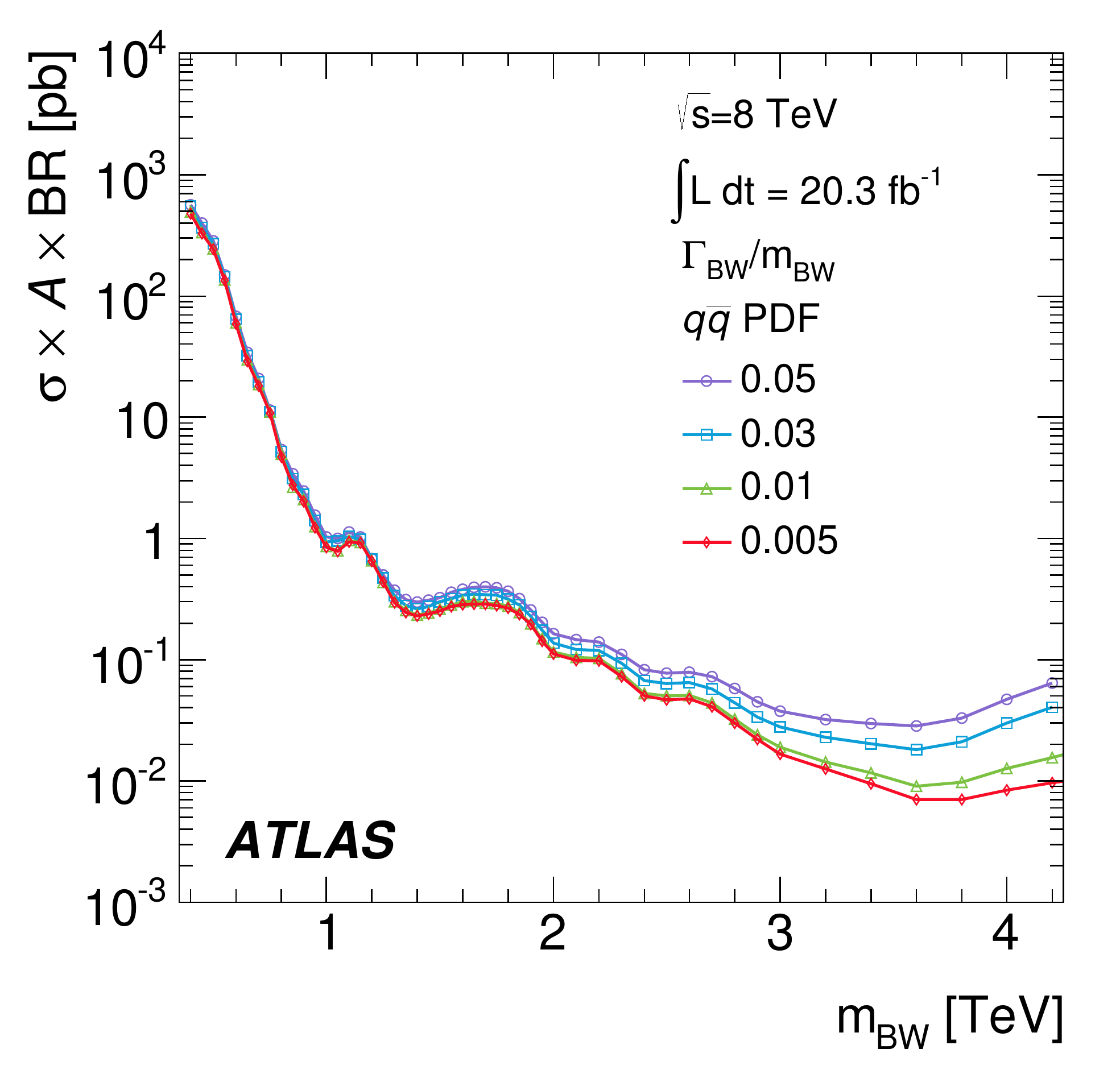}
  \caption{The 95\%\ C.L. upper limits on $\sigma\times {\cal A}$ for a Breit-Wigner narrow resonance
           produced by a $q\bar{q}$ initial state
           decaying to dijets and convolved with PDF effects, dijet mass acceptance, parton shower and nonperturbative effects and detector resolution,
           as a function of the mean mass, $m_{\text{BW}}$, for different values of intrinsic width over mass ($\Gamma_{\text{BW}}/m_{\text{BW}}$),
           taking into account both statistical and systematic uncertainties.}
  \label{fig:BWQQbar}
\end{figure}

The resulting limits on $\sigma\times {\cal A}$ for
the Gaussian template shape are shown in Fig.~\ref{fig:modindeplims}.
Limits resulting from the convolution of Breit-Wigner signals of different intrinsic widths ($\Gamma_{\rm BW}$) with the appropriate parton distribution function, parton shower, nonperturbative effects and detector resolution are shown in Figs.~\ref{fig:BWGG} and~\ref{fig:BWQQbar}.
For the initial Breit-Wigner signal the following nonrelativistic function was chosen:
\[
f(x,\mu,\Gamma)=\frac{1}{2 \pi} \frac{\Gamma}{(x-\mu)^2+(\Gamma^2/4)}
\],
where $\mu$ and $\Gamma$ are the mass and the width of the resonance.
The use of a relativistic Breit-Wigner signal for the resonance line shape may
lead to different limits than the ones derived using the nonrelativistic approximation above.
Parton showers and nonperturbative effects have been simulated using \Herwig 2.6.3,
which gives a more conservative limit with respect to what is obtained from \Pythia{}.

The difference in shapes between the two Breit-Wigner limits is a result of the much larger low-mass tails
resulting from the $gg$ parton luminosity, which becomes especially pronounced at high masses.
The convolution with parton shower and nonperturbative effects enhances this effect further.

For sufficiently narrow resonances, these results may be used to set limits on NP
models beyond those considered in the current studies,
as described in detail in Appendix~\ref{sec:genericNPlimits}.

It should be noted that these limits will be conservative at high masses with respect
to the limits obtained with full benchmark templates. This is due to the simplifying assumptions
made in their derivation, in particular from the use of a nonrelativistic and
mass-independent Breit-Wigner shape.

Gaussian limits should be used when tails from PDF and nonperturbative
effects can be safely truncated or neglected.
Otherwise, convolved Breit-Wigner signals would be more reliable.

In the case of the Gaussian limits, the signal
distribution after applying the
kinematic selection criteria on $y^*$, $\mjj$  and $\eta$ of the
leading jets (Sec.~\ref{sec:jetCalibration}) should approach a Gaussian distribution.
The acceptance should include the jet reconstruction efficiency (100\% for the current
analysis and detector conditions, since inefficiencies due to calorimeter problems are corrected for in data)
and the efficiency with respect to the kinematic selection above. NP models with a width smaller than 5\% should be compared to
the results with width equal to the experimental resolution only (see Appendix~\ref{sub:dijetmassresolution}).
For models with a larger width after detector effects, the limit that best matches their width should be used.

\clearpage

\section{Conclusions}

In the 2012 running of the ATLAS experiment at the LHC, the collision energy was raised
from 7~TeV to 8~TeV, accompanied by a more than fourfold increase in integrated luminosity.
The higher energy, and the associated rise in parton luminosity for high masses,
have increased the sensitivity of the search and its mass reach for various model hypotheses.
In addition, novel trigger techniques have been employed to extend the search to
low dijet masses. The data sample used in the current analysis consists of
\integLumi\ of $pp$ collision data at $\sqrt{s} = 8$~TeV, and the resulting
dijet mass distribution extends from 250~GeV to approximately $\HighestDijetMass$~TeV.

No resonancelike features are observed in the dijet mass spectrum.
This analysis places limits on the cross section times
acceptance at the 95\% credibility level on the mass or energy scale
of a variety of hypotheses for physics phenomena beyond the Standard Model.

To illustrate the typical increases in sensitivity to new phenomena
at the LHC up to the end of 2012 running, Table~\ref{tab:complims} shows
the history of expected limits from ATLAS studies using dijet resonance
analysis of two benchmark models, excited quarks and color-octet scalars.
The limits set by this analysis on excited quarks, color-octet scalars,
heavy $\Wprime$ bosons, chiral $\Wstar$ bosons, and quantum black holes,
are summarized in Table~\ref{tab:Summary}.

\begin{table}[h!]
\centering {
\begin{tabular}{lcc}
\hline
Model and final state  \quad \quad &  \multicolumn{2}{c}{95\%\ C.L. Limits [TeV]} \\
            &  expected  &  observed\\ \hline
$\qstar \rightarrow qg$    &   \MjjLimitExpectedqstar   & \MjjLimitqstar  \\
$s8 \rightarrow gg$    &    \MjjLimitExpecteds8     & \MjjLimits8  \\
$W'\rightarrow q\bar{q}'$  &   \MjjLimitExpectedWprime  & \MjjLimitWprime \\
Leptophobic $W^*\rightarrow q\bar{q}'$  &   \MjjLimitExpectedWstar         & \MjjLimitWstar  \\
Leptophilic $W^*\rightarrow q\bar{q}'$  &   \MjjLimitExpectedW1star         & \MjjLimitW1star  \\
\textsc{Qbh} black holes &   \MjjLimitExpectedQBH           & \MjjLimitQBH  \\
($q$ and $g$ decays only)  & & \\
\textsc{BlackMax} black holes&   \MjjLimitExpectedBlackMax      & \MjjLimitBlackMax \\
(all decays)  & & \\
\hline
\end{tabular}
}
\caption {The 95\% C.L. lower limits on the masses and energy scales
of the models examined in this study.  All limit analyses are Bayesian,
with statistical and systematic uncertainties included.
}
\label{tab:Summary}
\end{table}

\begin{table}[!htb]
\begin{center}
\begin{tabular}{lllll}
\hline
  $\sqrt{s}$  &  $\mathcal{L}$   &  Citation  &  $\qstar$ [TeV] &  $s8$ [TeV] \\
\hline
  7 TeV       &  36 \invpb{}  & \cite{Aad:2011aj}          & 2.07 & ~~-    \\
  7 TeV       &  1.0 \invfb{} & \cite{ATLAS:Res2011}       & 2.81 & 1.77 \\
  7 TeV       &  4.8 \invfb{} & \cite{ATLAS:2012pu}        & 2.94 & 1.97 \\
  8 TeV       & \integLumi  &  current & \MjjLimitExpectedqstar & \MjjLimitExpecteds8 \\
\hline
\end{tabular}
\end{center}
\caption{ATLAS previous and current expected 95\% C.L. upper limits
  [TeV] on excited quarks and color-octet scalars.
}
\label{tab:complims}
\end{table}

\bigskip
\section*{Acknowledgments}

We thank CERN for the very successful operation of the LHC, as well as the
support staff from our institutions without whom ATLAS could not be
operated efficiently.

We acknowledge the support of ANPCyT, Argentina; YerPhI, Armenia; ARC,
Australia; BMWF and FWF, Austria; ANAS, Azerbaijan; SSTC, Belarus; CNPq and FAPESP,
Brazil; NSERC, NRC and CFI, Canada; CERN; CONICYT, Chile; CAS, MOST and NSFC,
China; COLCIENCIAS, Colombia; MSMT CR, MPO CR and VSC CR, Czech Republic;
DNRF, DNSRC and Lundbeck Foundation, Denmark; EPLANET, ERC and NSRF, European Union;
IN2P3-CNRS, CEA-DSM/IRFU, France; GNSF, Georgia; BMBF, DFG, HGF, MPG and AvH
Foundation, Germany; GSRT and NSRF, Greece; RGC, Hong Kong SAR, China; ISF, MINERVA, GIF, I-CORE and Benoziyo Center,
Israel; INFN, Italy; MEXT and JSPS, Japan; CNRST, Morocco; FOM and NWO,
Netherlands; BRF and RCN, Norway; MNiSW and NCN, Poland; GRICES and FCT, Portugal; MNE/IFA, Romania; MES of Russia and ROSATOM, Russian Federation; JINR; MSTD,
Serbia; MSSR, Slovakia; ARRS and MIZ\v{S}, Slovenia; DST/NRF, South Africa;
MINECO, Spain; SRC and Wallenberg Foundation, Sweden; SER, SNSF and Cantons of
Bern and Geneva, Switzerland; NSC, Taiwan; TAEK, Turkey; STFC, the Royal
Society and Leverhulme Trust, United Kingdom; DOE and NSF, United States of
America.

The crucial computing support from all WLCG partners is acknowledged
gratefully, in particular from CERN and the ATLAS Tier-1 facilities at
TRIUMF (Canada), NDGF (Denmark, Norway, Sweden), CC-IN2P3 (France),
KIT/GridKA (Germany), INFN-CNAF (Italy), NL-T1 (Netherlands), PIC (Spain),
ASGC (Taiwan), RAL (UK) and BNL (USA) and in the Tier-2 facilities
worldwide.

\bigskip

\appendix
\section{Suggested procedure for setting limits on generic NP models}
\label{sec:genericNPlimits}

\subsection{Setting limits for NP models with a Gaussian shape}
\label{sec:GaussianLimit}

The following detailed procedure is appropriate for setting limits
involving resonances that are approximately Gaussian near the core,
and with tails that are much smaller than the background.
The results of Fig.~\ref{fig:modindeplims} are provided in
tables on HepData.

(1) Generate an MC sample of a hypothetical new particle with mass
set to $M$. nonperturbative effects should be included in the event generation.
Apply the kinematic selection on the parton $\eta$, \pT, and $|y^*|$ used in this analysis,
as in Sec.~\ref{sec:jetCalibration}.

(2) Smear the signal mass distribution to reflect the detector resolution.
The smearing factors derived from full ATLAS simulation of QCD dijet events
can be taken from Fig.~\ref{fig:massResolution}.

(3) Since a Gaussian signal shape has been assumed in determining the limits,
any long tails in the reconstructed $\mjj$\ should be removed in the sample under study.
The recommendation (based on optimization using $q^\ast$ templates) is to retain
events with $\mjj$\ between 0.8$M$ and 1.2$M$.
The mean mass, $m$, should be recalculated for this truncated signal.

(4) The fraction of MC events surviving the first four steps determines
the modified acceptance, ${\cal A}$.

(5) From the table in HepData,
select $m_{\text{G}}$ so that $m_{\text{G}}=m$.
If the exact value of $m$ is not among the listed values of $m_{\text{G}}$,
check the limit for the two values of $m_{\text{G}}$ that are directly above and below
$m$, and use the larger of the two limits to be conservative.

(6) To retain enough of the information in the full signal template, and at the
same time reject tails that would invalidate the Gaussian approximation, the
following truncation procedure is recommended.
For this mass point, choose a value of $\sigma_{\text{G}}/m_{\text{G}}$
such that the region within $\pm 2\sigma_{\text{G}}$ is well contained in the (truncated) mass range.
For the $\qstar$ case a good choice is $\sigma_{\text{G}}$ =
(1.2$M$-0.8$M$)/5 so that 95\% of the Gaussian spans $4 \times(0.4/5)M$.
Use this value to pick the closest $\sigma_{\text{G}}$/$m_{\text{G}}$ value,
rounded up to be conservative.

(7) Compare the tabulated 95\% C.L. upper limit corresponding
to the chosen $m_{\text{G}}$ and $\sigma_{\text{G}}/m_{\text{G}}$
values to the $\sigma\times {\cal A}$ obtained from the theoretical
cross section of the model multiplied by the acceptance defined
in step (4) above and taking into account its branching ratio into dijets.

\subsection{Setting limits for NP models with a Breit-Wigner shape, accounting for PDF effects}
\label{sec:BWLimit}

The following detailed procedure is appropriate for setting limits
involving resonances that approximate a Breit-Wigner (BW) shape and
extend with a low-mass tail due to effects of parton luminosity.
For signals that are very narrow or whose tails deviate significantly
from a BW, a truncation of the signal template suggested in the
Gaussian limits in Sec.~\ref{sec:GaussianLimit} might be more appropriate.
The results of Figs.~\ref{fig:BWGG} and~\ref{fig:BWQQbar} are provided in Tables I and II on HepData.

(1) Generate a hypothetical new particle, with mass set to $M$
and intrinsic width $\Gamma$. As the PDF used to obtain those limits
is CT10, the same choice is recommended for the event generation.
nonperturbative effects should be simulated after the hard scattering.

(2) Smear the signal mass distribution to reflect the detector resolution.
The smearing factors for the dijet mass are derived from full
ATLAS simulation and can be taken from Fig.~\ref{fig:massResolution}.

(3) The kinematic selection detailed in Sec.~\ref{sec:jetCalibration}
should be applied to the simulated events. It should be checked at this point
that the shape of the template after the $y^*<0.6$ cut does not change significantly.
For example, in a simple model with a flat distribution for the cosine of the polar angle of the jets
in the rest frame of the resonance ($\cos\theta^*$)
that decays into two back-to-back jets ($y_{\text{lead}} \approx - y_{\text{sublead}}$),
a cut on $|y^*|=0.5*|y_{\text{lead}}-y_{\text{sublead}}|<0.6$ imposes a cut on the unboosted rapidity distribution ($|y_{\text{lead,sublead}}|<0.6$).
In the mass ranges investigated, this corresponds to a
more stringent constraint than the $\eta_{\rm BW}<2.8$
acceptance correction, leading to an acceptance
of $\sim 0.5$ that does not depend on dijet mass.
Deviations from a flat acceptance of up to 20\% can be observed in the tails of models
with different angular distributions ($q^*, s8, W'$).

(4) The fraction of generated events surviving the first three steps determines
the signal acceptance, ${\cal A}$.

(5) From the tables available in HepData,
select the one corresponding to the production mode for the new resonance
(gluon-gluon, gluon-quark, quark-quark or quark-antiquark)
as the parton luminosities and hence the signal shapes differ.

(6) Compare the tabulated 95\% C.L. upper limit corresponding
to the chosen $M$ and $\Gamma/M$
values to the $\sigma\times {\cal A}$ obtained from the theoretical
cross section of the model, multiplied by the acceptance defined in step (3) above
and taking into account its branching ratio into dijets.

If the exact values of $M$ and $\Gamma/M$ are not among
the listed values of $m_{\text{BW}}$ and  $\Gamma_{\text{BW}}/m_{\text{BW}}$,
check the limit for the two values of $m_{\text{BW}}$ that are directly above and below
$M$, and use the more conservative of the two limits.

\section{Dijet mass resolution}
\label{sub:dijetmassresolution}

\begin{figure}[!htb]
  \centering \includegraphics[width=0.45\textwidth]{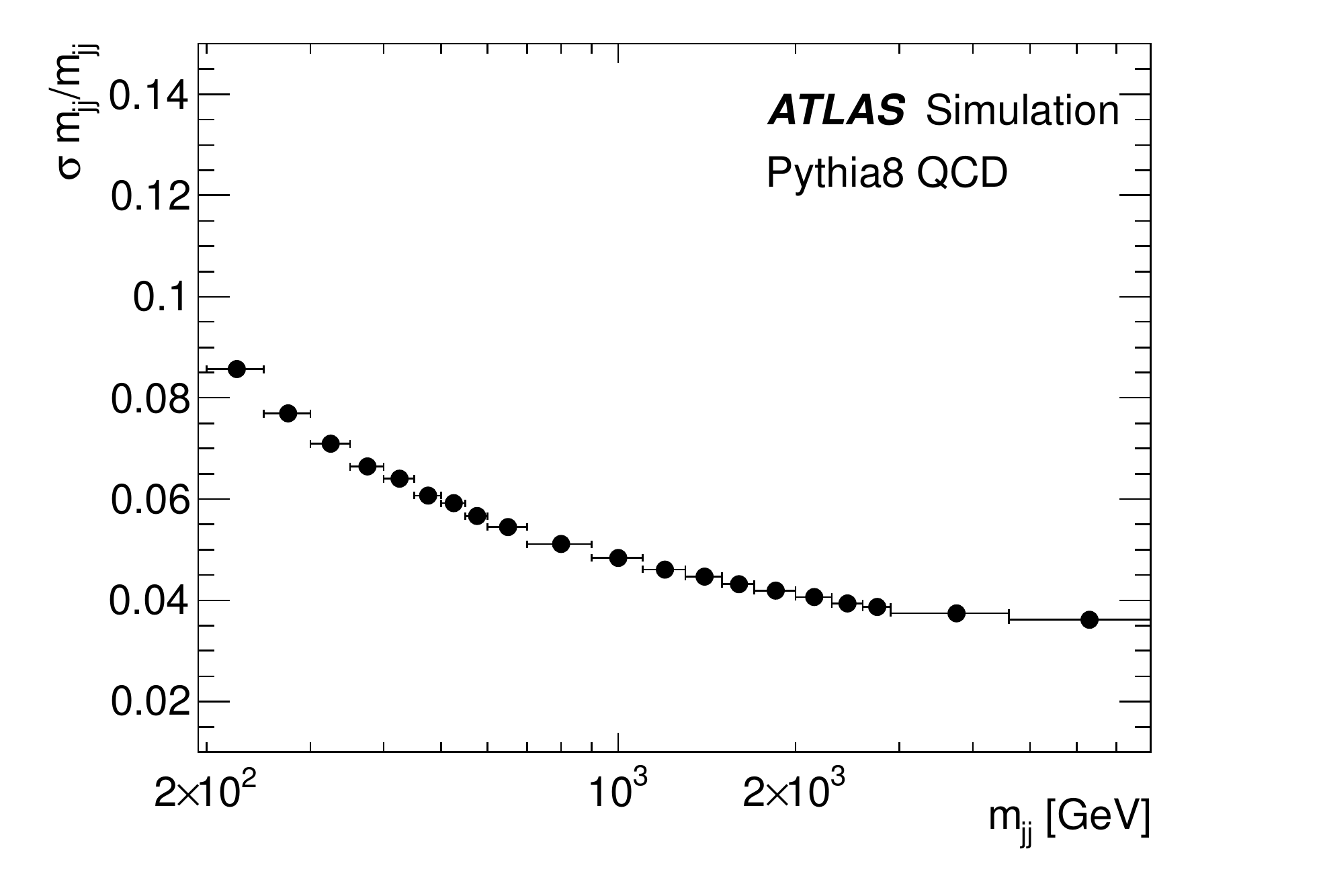}
  \caption{Dijet mass resolution obtained from fully simulated \Pythia\ QCD Monte Carlo
\Pythia{} 8.175~\cite{Pythia8}, with the AU2 tune obtained from ATLAS data~\cite{ATL-PHYS-PUB-2012-003}.
The dijet mass resolution is interpolated linearly between the bin centers. }
  \label{fig:massResolution}
\end{figure}

The dijet mass resolution in Fig.~\ref{fig:massResolution} is derived from fully simulated QCD Monte Carlo, generated with
\Pythia{} 8.175~\cite{Pythia8}, using the AU2 tune obtained from ATLAS data~\cite{ATL-PHYS-PUB-2012-003}
using the analysis selection detailed in Sec.~\ref{sec:jetCalibration}.
The dijet mass resolution $\sigma_{m_{jj}} / m_{jj}$ is \MjjResolutionLow{}
at $m_{jj}\simeq$250 GeV,
falls to 4\% at 2 TeV, and approaches
\MjjResolutionHigh{} at $m_{jj}$
of 3 TeV and above, and it is interpolated linearly between
the bin centers.

\section{Signal template shapes}
\label{sub:signalTemplateShapes}

\begin{figure}[!htb]
  \centering \includegraphics[width=0.45\textwidth]{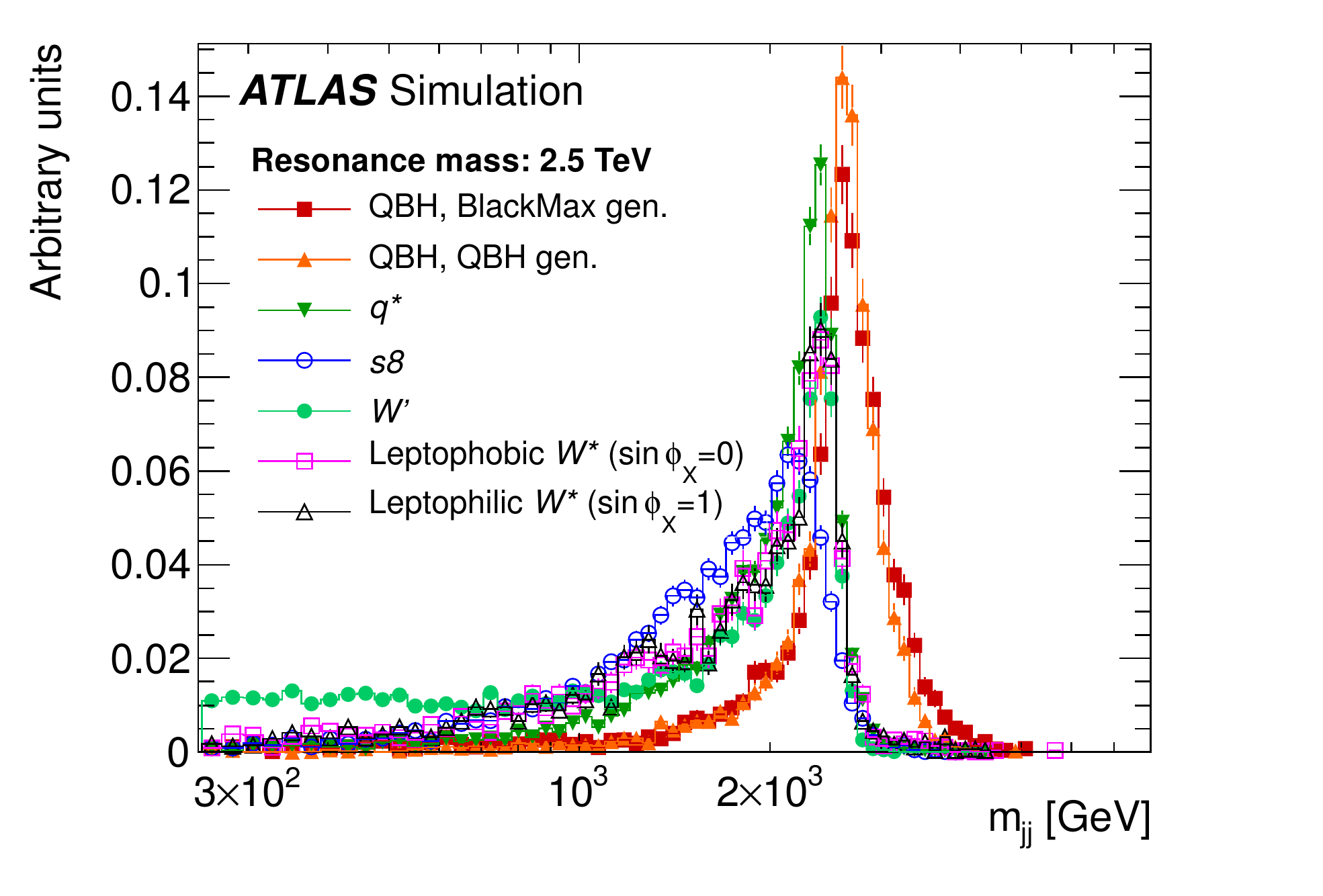}
  \caption{Dijet invariant mass for models corresponding to a resonance mass of 2.5 TeV. All distributions are normalized to the same area. }
  \label{fig:templateShapes}
\end{figure}

For ease of comparison of the shapes of different signals used in this paper, the various signal template shapes are overlaid in Fig.~\ref{fig:templateShapes} for the mass point of 2.5 TeV, after normalizing to the same area.

\clearpage

\onecolumngrid
\clearpage
\begin{flushleft}
{\Large The ATLAS Collaboration}

\bigskip

G.~Aad$^{\rm 84}$,
B.~Abbott$^{\rm 112}$,
J.~Abdallah$^{\rm 152}$,
S.~Abdel~Khalek$^{\rm 116}$,
O.~Abdinov$^{\rm 11}$,
R.~Aben$^{\rm 106}$,
B.~Abi$^{\rm 113}$,
M.~Abolins$^{\rm 89}$,
O.S.~AbouZeid$^{\rm 159}$,
H.~Abramowicz$^{\rm 154}$
% ...
\bigskip
\\
% ...
$^{11}$ Institute of Physics, Azerbaijan Academy of Sciences, Baku, Azerbaijan\\
% ...
$^{84}$ CPPM, Aix-Marseille Universit{\'e} and CNRS/IN2P3, Marseille, France\\
% ...
$^{89}$ Department of Physics and Astronomy, Michigan State University, East Lansing MI, United States of America\\
% ...
$^{106}$ Nikhef National Institute for Subatomic Physics and University of Amsterdam, Amsterdam, Netherlands\\
% ...
$^{112}$ Homer L. Dodge Department of Physics and Astronomy, University of Oklahoma, Norman OK, United States of America\\
$^{113}$ Department of Physics, Oklahoma State University, Stillwater OK, United States of America\\
% ...
$^{116}$ LAL, Universit{\'e} Paris-Sud and CNRS/IN2P3, Orsay, France\\
% ...
$^{152}$ Institute of Physics, Academia Sinica, Taipei, Taiwan\\
% ...
$^{154}$ Raymond and Beverly Sackler School of Physics and Astronomy, Tel Aviv University, Tel Aviv, Israel\\
% ...
$^{159}$ Department of Physics, University of Toronto, Toronto ON, Canada\\
% ...
\end{flushleft}


\end{document}